\documentclass[aps,prd,twocolumn,superscriptaddress,showpacs,floatfix,10pt]{revtex4}
\usepackage{graphicx}
\usepackage{subfigure}
\usepackage{epsfig}

\begin{document}

\title{Bounded modes to the rescue of optical transmission}

\author{Micha\"{e}l Sarrazin}
\email{michael.sarrazin@fundp.ac.be} \affiliation{Laboratoire de
Physique du Solide, Facult\'es Universitaires Notre-Dame de la Paix,
\\61 rue de Bruxelles, B-5000 Namur, Belgium}

\author{Jean Pol Vigneron}
\email{jean-pol.vigneron@fundp.ac.be} \affiliation{Laboratoire de
Physique du Solide, Facult\'es Universitaires Notre-Dame de la Paix,
\\61 rue de Bruxelles, B-5000 Namur, Belgium}

\begin{abstract}
This paper presents a brief survey of the evolution of knowledge
about diffraction gratings. After recalling some basic facts,
historically and physically, we introduce the concept of Wood
anomalies. Next, we present some recent works in order to
introduce the role of bounded modes in transmission gratings. The
consequences of these recent results are then introduced.
\\ \textit{This paper is a secondary publication, published
in Europhysics News (EPN \textbf{38}, 3 (2007) 27-31). In the
present version, some additional notes have been added with
related references.}
\end{abstract}


%

\maketitle

In the beginning of the 19$^{th}$ century, T. Young and J. Fraunhofer built
the first optical diffraction gratings and showed the role of optical
diffraction in their behaviour. Since that time, diffraction gratings have
been used in a broad range of technological applications, from the earlier
accurate optical spectrometers to the recent integrated optical devices.
They are also used in a broad range of wavelength from X-ray to microwaves.
In this context, diffraction gratings have been the cornerstones of many
publications for almost two centuries. Surprisingly, nowadays the subject is
far from being exhausted.

\begin{figure}[b]
\centerline{
\includegraphics[width=8 cm]{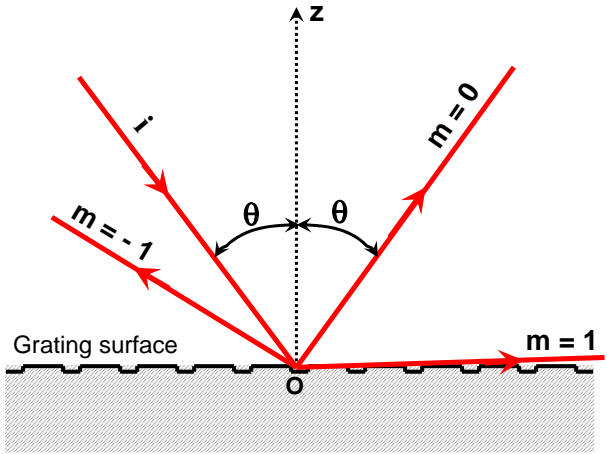}}
\caption{A reflecting optical grating. $i$ is the incident light
beam. The light beams labelled $m$ are some of the diffracted
orders and $\theta$ is the angle of incidence.} \label{fig1}
\end{figure}

Every student knows that a grating can spread a white light beam from an
incandescent lamp into a continuous spectrum of colors. For a grating made
with a one-dimensional lattice of thick-thin wires, this ''rainbow'' is
duplicated many times, each one corresponding to a diffraction order $m$
(See Fig.1 for instance). These basic behaviours of diffraction gratings can
be easily understood by the conservation of light momentum such as

\begin{equation}
\mathbf{k}_s=\mathbf{k}_i+\mathbf{K}  \label{1}
\end{equation}
where $\mathbf{k}_s$ is the momentum of the scattered light, $\mathbf{k}_i$
is the momentum of the incident light, and $\mathbf{K}$ is a vector of the
reciprocal lattice of the grating such that $\mathbf{K\cdot a=}2\pi m$ ($%
m\in Z$), where $\mathbf{a}$ is the basis vector of the lattice. The general
behaviour of optical grating results from this basic assumption. For
instance, in Fig.1, one considers a reflecting grating. An incident light $i$
is then scattered into three diffraction orders. The zero order ($m=0$) is
known as the specular beam. In Fig.1, two other orders are considered, $m=1$
and $m=-1$.

\section{Wood Anomalies}

In 1902, R. W. Wood observed some unexpected patterns in the spectrum of
light resolved by optical diffraction gratings [1]. The Wood spectrum
presented many unusual rapid variations of its intensity in certain narrow
wavelengths bands. As a consequence, these effects unexplained by ordinary
grating theory were named ''Wood anomalies''.

In 1907, Lord Rayleigh proposed an explanation of those anomalies [2].
Considering for instance the case of the Fig.1, one notes that the order
becomes tangent to the surface grating before disappearing. In such a case,
the light momentum component along the $Oz$ axis of a diffracted order $m$,
i.e. $\mathbf{k}_{z,m}$, becomes imaginary right after having been
cancelled. Then, the diffracted order $m=1$ becomes evanescent
(non-homogeneous order). When $\mathbf{k}_{z,m}$ is real, the diffracted
order is free to propagate along the $Oz$ axis (homogeneous order). In the
present case, as the order $m=1$ turns to be non-homogeneous, its energy
will be redistributed over the other orders. As a consequence, the
diffracted beam intensity of the specular beam ($m=0$), for instance,
increases just as the diffracted order $m=1$ vanishes. For an incidence $%
\theta $ (see Fig.1), a diffracted order $m$ becomes non-homogeneous for
wavelengths greater than a specific value, the Rayleigh wavelength, for
which an anomaly occurs. One then talks about a Rayleigh anomaly.

Around 1938, U. Fano proposed another explanation [3], where the anomalies
are related to a resonance effect. Such a resonance comes from a coupling
between an eigenmode of the grating and a non-homogeneous diffraction order
(one also talks about a resonant diffraction order), i.e. after it vanishes
for wavelengths greater than the corresponding Rayleigh wavelength.

From an experimental point of view, gratings can then show some very complex
behaviour often hard to explain in details without the use of Maxwell's
equations. In addition, solving Maxwell's equations in order to study the
electromagnetic diffraction quickly becomes a difficult problem, even with
simple geometries. Except for some simple and powerful analytical works,
such as those from Lord Rayleigh or U. Fano, it will be necessary to wait
until the Sixties and the beginning of computer calculation to observe
significant progress in the Wood anomalies comprehension. For instance, one
can underline the first fundamental numerical results of Hessel and Oliner
in 1965 [4], and Maystre and Nevi\`{e}re in 1977 [5,6]. Hessel and Oliner
presented a wide study of the eigenmode resonance role in Wood anomalies
[4]. They have shown that depending on the type of periodic structure, the
two kinds of anomalies i.e., Rayleigh anomalies or resonant anomalies may
occur separately or are almost superimposed.

Maystre and Nevi\`{e}re studied specific cases of such resonant anomalies
[5,6]. For instance, they presented a wide study about ''plasmon
anomalies'', which occurs when the surface plasmons of a metallic grating
are excited [5]. They also considered an anomaly that appears when a
dielectric coating is deposited on a metallic grating, and corresponds to
guided modes resonances in the dielectric layer [6].

It is important to underline that, in some recent publications, some authors
improperly reduce the Wood anomalies to the Rayleigh anomalies only. It is
important to recall that the Wood anomaly concept covers the resonant
anomalies (Fano anomalies) and the Rayleigh anomalies at the same time. In
addition, the resonant part of the phenomenon can imply many kinds of
resonances and not only the surface plasmons resonances.

Amazingly, thirty years after the pioneer works, in the early 21$^{st}$
century, the experimental and theoretical interest in these subjects has not
dried up.

\section{Optical Transmission in a Nanostructured Slab}

In 1998, Ebbesen \textit{et al} [7] reported the amazing properties of the
optical transmission of some optical gratings. These were made by drilling
cylindrical holes in a thin metallic layer along a bidimensional lattice.
The metallic layer was deposited on a glass substrate (see Fig.2). These
experiments renewed the motivation for investigating metallic gratings, in
particular those with a two-dimensional lattice. The most attractive
characteristic of their results was the peculiar wavelength dependence of
the transmission (see Fig.3). The latter was defined as the ratio between
the energy of the specular transmitted beam ($m=0$) only and the energy of
the incident beam. After these first experimental observations, the role of
the thin metallic film surface plasmons (SPs) was put forward in order to
explain the peculiar wavelength dependence of the transmission [7].

\begin{figure}[t]
\centerline{
\includegraphics[width=6 cm]{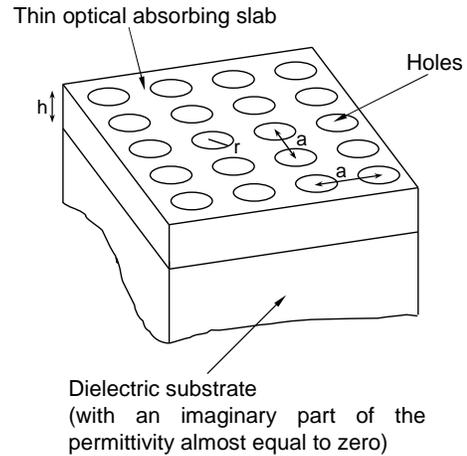}}
\caption{The typical bidimensional array of subwavelength holes.}
\label{fig2}
\end{figure}

A surface plasmon on a plane surface is a non-radiative electromagnetic
mode, associated with a collective excitation of electrons at the interface
between a conductor and an insulator. In this way, a surface plasmon cannot
be excited directly by light, and it cannot decay spontaneously into
photons. The non-radiative nature of SPs is due to the fact that interaction
between light and SPs cannot simultaneously satisfy energy and momentum
conservation laws. Moreover, the momentum conservation requirement can be
achieved by roughening or corrugating the metal surface, for instance. It is
exactly the situation in those devices. In this case, the transmission peaks
were interpreted as SPs resonances [7]. Surprisingly, this approach seems to
have been suggested independently of the knowledge of some earlier works
about the reflecting one-dimensional gratings. Then, the first attempt to
understand the Ebbesen experiments has missed some important earlier results
of the optical gratings study.

\begin{figure}[t]
\centerline{
\includegraphics[width=8 cm]{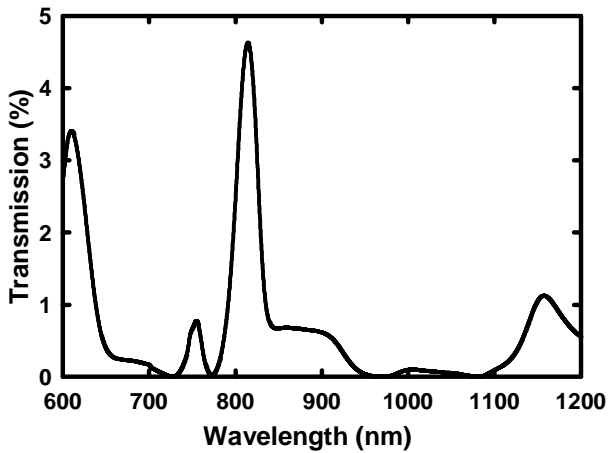}}
\caption{Computed optical transmission of a thin gold film ($200$
nm) deposited on glass. $100$ nm is set for the radius of each
hole and the lattice parameter is taken to be $a$ $=$ $700$ nm.}
\label{fig3}
\end{figure}

In this way, in these first interpretations, the exact role of SPs was not
clearly assessed, and many questions remained to clarify completely the
scattering processes involved in these experiments.

Note that, some authors developed other explanations as an alternative to
the SPs model. For instance, it was suggested that the transmission pattern
could also be described as resonances of electromagnetic cavity modes (here
the cavities are the holes in the studied gratings) [8]. More recently, T.
Thio and H. Lezec also suggested a diffracted evanescent wave model [9].

In addition, some doubts against the SPs hypothesis relied on some
experimental results about subwavelength hole arrays made in non-metallic
films. Though SPs cannot exist, the transmission pattern was found to be
very similar to that obtained with a metallic film [9]. Surprisingly, the
very fact that the typical transmission pattern can be observed even in
non-metallic systems convinced some authors to fully reject the SPs
hypothesis and rather consider models involving nonresonant evanescent waves
diffraction. To our knowledge, this point of view is not supported by recent
results. In fact, the SPs must be replaced by other kinds of eigenmodes in
the nonmetallic cases.

As a consequence, despite alternative theoretical interpretations, a large
experimental consensus, many theoretical results, tend to prove that the SPs
interpretation was correct in the metallic films cases, and must be extended
to a bounded mode interpretation in the most general cases [10-12]. It is
only the transposition of the 1960's results about one-dimensional
reflection gratings to bidimensional transmission gratings.

Let us clarify this.

\section{Bounded Modes and Transmission}

Basically, the gratings considered here are simply drilled slabs. In this
case, a bounded mode, i.e. an eigenmode of the grating, is an
electromagnetic field configuration such that the field is trapped along the
slab. The mode can propagate along the slab, but not elsewhere. The
amplitude of the field decays then, exponentially from each side of the
slab. Typically, two amplitude patterns exist, which correspond to symmetric
and antisymmetric modes. As the thickness of the slab increases, in some
cases, the modes can refer to pure surface modes (Fig.4), i.e. both slab
interfaces are decoupled. Surface plasmons are such modes for instance.

\begin{figure}[t]
\centerline{
\includegraphics[width=8 cm]{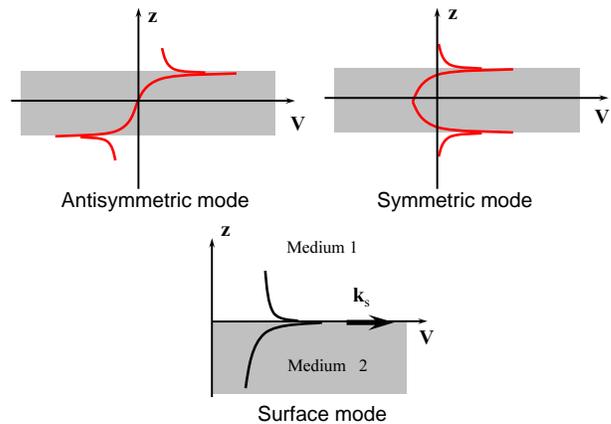}}
\caption{Diagrams of slab modes. The electrostatic potential $V$
of modes is plotted as a function of the location in the slab.
Above: antisymmetric and symmetric modes. Below: pure surface
mode, with each interface decoupled from each other.} \label{fig4}
\end{figure}

As previously suggested, bounded modes cannot be excited directly
by incident light or decay spontaneously into photons. Moreover,
the coupling between the outside electromagnetic field and the
bounded modes is allowed by the use of rough or corrugated
surface, as in the present case. The eigenmodes, and their
coupling with outside light, can be described mathematically as
follow. As in quantum mechanics, a scattering problem can be
treated in electromagnetism via the use of the scattering matrix
formalism ($S$ matrix). In such a representation, it is assumed
that the electromagnetic field can be described by two
super-vectors $\left| F_{in}\right\rangle $ and $\left|
F_{scat}\right\rangle $ (with the use of ''ket'' representation),
which correspond to the incident electromagnetic field that lit
the studied device and to the field scattered by the device. For
instance, each component of these vectors corresponds to a
specific diffraction order. Then, incident and scattered field are
linked via the scattering matrix defined as

\begin{equation}
\left| F_{scat}\right\rangle =S(\lambda )\left| F_{in}\right\rangle
\label{2}
\end{equation}
so that $S(\lambda )$ contains all the physical information about the
studied device, such as a diffraction grating. In addition, it can be shown
that $S(\lambda )$ can be explicitly defined from the Maxwell's equations.
For instance, the simulations in our papers are based on such a method,
which combines scattering matrix formalism with a plane wave representation
of the fields. This technique provides a computation scheme for the
amplitude and polarization (s or p) of reflected and transmitted fields in
any diffracted order. A fundamental key to the understanding of diffraction
grating phenomenology is to define the eigenmodes of the studied device.
Eigenmodes then obey the homogeneous equation
\begin{equation}
S^{-1}(\lambda _p)\left| F_{eigen}\right\rangle =0  \label{3}
\end{equation}
in such a way that the eigenmodes can also be defined by the poles $\lambda
_p$ of the scattering matrix. Then, the poles are the solutions of the
equation
\begin{equation}
\det \left\{ S^{-1}(\lambda _p)\right\} =0  \label{4}
\end{equation}
If we extract the singular part of $S(\lambda )$ corresponding to the
eigenmodes of the structure, we can write $S(\lambda )$ in an analytical
form as
\begin{equation}
S(\lambda )=\sum_p\frac{R_p}{\lambda -\lambda _p}+S_h(\lambda )  \label{5}
\end{equation}
This is a generalized Laurent series, where $R_p$ are the residues
associated with each pole $\lambda _p$. One notes that $\lambda _p=\lambda
_p^R+i\lambda _p^I$ are complex numbers, and that the imaginary part $%
\lambda _p^I$ can be linked to lifetimes of the eigenmodes. The divergent
behaviour of the fractional terms underlines the resonant character of the
eigenmodes (one talks then about the resonant terms of the $S$ matrix). $%
S_h(\lambda )$ is the holomorphic part of $S(\lambda )$ which corresponds to
purely non-resonant processes. Ideally, in the vicinity of a specific pole,
the holomorphic part can be assumed to be almost constant. Then, as the
amplitude $s(\lambda )$ of a specific diffraction order is related to an
element of the scattering matrix, the intensity $I\propto \left| s(\lambda
)\right| ^2$ of an order can be easily derived. One then obtains
\begin{equation}
I_F=\widetilde{I_0}\frac{\left( \lambda -\lambda _z^R\right) ^2+\lambda
_z^{I\ \ 2}}{\left( \lambda -\lambda _p^R\right) ^2+\lambda _p^{I\ \ 2}}
\label{6}
\end{equation}
where $\lambda _z=\lambda _z^R+i\lambda _z^I$ is the complex zero of $%
s(\lambda )$ in the vicinity of $\lambda _p$ in the complex plan.

\begin{figure}[t]
\centerline{
\includegraphics[width=8 cm]{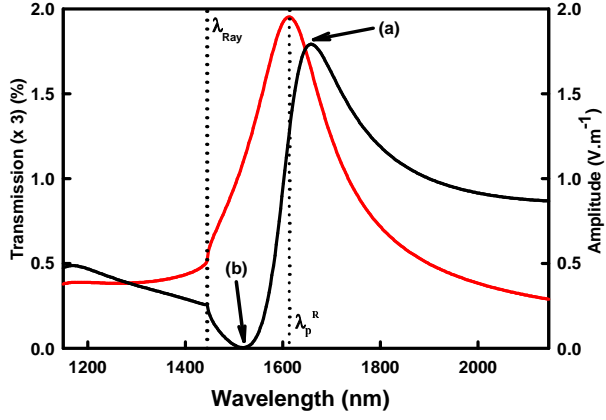}}
\caption{Computed optical transmission (black curve) and amplitude
of the related resonant order (red curve) of a thin fictitious
film (the permittivity is equal to $-25+i1$) deposited on glass.
$500$ nm is set for the radius of each hole and the lattice parameter is taken to be $a$ $=$ $%
1000$ nm. A discontinuity appears at the Rayleigh wavelength
$\lambda _{Ray}$.} \label{fig5}
\end{figure}

This function leads to a typical asymmetrical profile, a Fano
profile, related to a Fano resonance. A typical Fano profile
appears in Fig.5 (black curve). Now, if the holomorphic part is
considered as negligible, one gets the following expression of the
amplitude
\begin{equation}
I_{BR}=I_0\frac 1{\left( \lambda -\lambda _p^R\right) ^2+\lambda _p^{I\ \ 2}}
\label{7}
\end{equation}
which gives a simple Lorentzian profile, related to a Breit-Wigner resonance
(see for instance the red curve of Fig.5). As shown in Fig.5, for a given
eigenmode (i.e. $\lambda _p$ imposed), the value of $\lambda $ for which $%
I_{BR}$ is a maximum (i.e. $\lambda =\lambda _p^R$) does not correspond to
the location of $I_F$ maximum or minimum (i.e. (a) or (b) in Fig.5). In
order to illustrate the physical mechanisms responsible for the behaviour
described by the scattering matrix formalism, we represent the corresponding
processes involved.

\begin{figure}[t]
\centerline{
\includegraphics[width=8 cm]{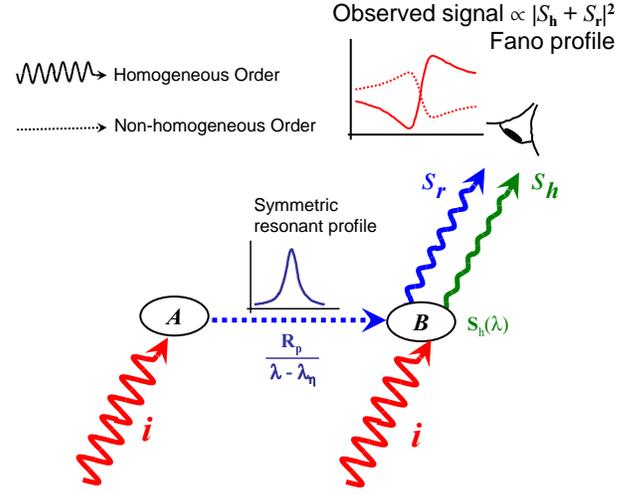}}
\caption{Diagrammatic representation of the processes responsible
for the behaviour of the transmission properties.} \label{fig6}
\end{figure}

In Fig. 6, circles $A$ and $B$ represent diffracting elements
e.g., holes. So an incident homogeneous wave $i$ (in red)
diffracts in $A$ and generates a nonhomogeneous diffraction order
(blue dashed line). Such an order is coupled with an eigenmode,
which is characterized by a complex wavelength $\lambda _p$. It
then becomes possible to excite this eigenmode, which leads to a
feedback reaction on the nonhomogeneous order (one can talk then
about a resonant order). This process is related to the resonant
term of the scattering matrix. In Fig.5, the red curve is the
electromagnetic field amplitude calculated for such a
non-homogeneous resonant order. The resonant diffraction order
diffracts then in $B$, and generates a contribution to a
homogenous diffraction order, in our case the specular beam. Thus
one can ideally expect to observe a resonant profile, i.e.
Lorentzian-like, for the amplitude $s_r$ of the homogenous
specular diffraction order that appears in $B$. Nevertheless, it
is also necessary to account for nonresonant diffraction processes
related to the holomorphic term $S_h(\lambda )$. So the incident
wave $i$, generates also a homogeneous order $s_h$ (in green),
which appears as another contribution to the specular beam. So, as
one observes the specular transmitted beam, one observes the sum
of two rates, i.e. $s_r+s_h$. So, the Fano profiles of the
transmission (but also of the reflection) result from a
superposition of resonant and nonresonant contributions to the
observed diffraction order [10-11]. As a result, in Fig.5 one can
observe the calculated transmission (black curve) in comparison
with the non-homogeneous resonant order amplitude (red curve).

\section{Consequences and Observations}

As shown in Fig.3, the observed optical transmission through subwavelength
hole arrays, exhibits a set of peaks and dips. Following the previously
described approach, we have shown in recent works that the transmission
spectrum is better depicted as a series of Fano profiles (see Fig.3 and
Fig.5) [10,13,14]. These recognizable line shapes result from the
interference of nonresonant transfers with resonant transfers, which involve
the eigenmodes film, and evanescent diffraction orders. We could then point
out that each transmission peak-dip pair is nothing other than a Fano
profile. The appearance of an assymetric Fano line shape does not
necessarily locate the peak or the dip at the eigenmode resonance (as shown
in Fig.5), contrary to what has been suggested in some earlier works [7,8].
However, we have shown that the existence of the eigenmodes is a condition
for the presence of the Fano line shapes. It can be noted that recent
results by Genet \textit{et al} confirm this description [11]. As a further
outcome of this work it became clear that this kind of transmission
spectrum, with its Fano line shapes, could also be obtained in a large
context and that the generic concept of eigenmodes must substitute that of
SPs only. These results implied two important ideas. First, according to the
contributions of resonant and nonresonant processes in the Fano profile,
eigenmodes can be associated with wavelengths closer to the peaks or to the
dips. Second, it is possible to obtain transmission curves similar to those
of metal films, by substituting SPs with guided modes or other types of
polaritons. Many examples can be found, including highly refractive
materials defining guided modes or ionic crystals in the restrahlen band
defining phonon polaritons (the restrahlen band of an ionic crystal is a
narrow domain of wavelength for which real part of the permittivity becomes
negative like in metals). For instance, we have reported simulations of a
device, which consists of arrays of subwavelength cylindrical holes in a
tungsten layer deposited on a glass substrate [13]. Tungsten becomes
dielectric on a restricted domain of wavelength in the range $240$-$920$ nm,
i.e., the real part of its permittivity becomes positive. So, whereas
plasmons cannot exist, we show that the transmission pattern is similar to
that obtained in the case of a metallic film. Indeed, it was also shown that
in this case, instead of SPs, guided modes are excited [13]. Nevertheless,
these modes give less intense fields than those achieved with SPs. In this
way the benefit for the transmission is less important than with SPs modes.
In the same way, it was also demonstrated that the transmission profile of a
chromium film, in the restricted wavelength domain ($1112$-$1292$ nm), where
the dielectric constant is positive, should involve eigenmodes, which are
not SPs or guided modes. In this case, SPs can be substituted by
Brewster-Zennek (BZ) modes [14]. Lastly it should be noted that in both
cases, these theoretical computations are supported by experimental results.

As a consequence, it appears that bounded modes can act as a mediator of the
optical transmission in bidimensional subwavelength hole arrays. Although it
appears that the mechanisms involved in these devices have been implicitly
known for a long time, it seemed important to us to recall it. Indeed, an
important engine of the technological development is also the improvement of
our theoretical knowledge of the phenomena with which we are confronted.

\textit{Additional notes :
\\- The theoretical results about
chromium films are corroborated by the experimental data of Ref.
[15].
\\- It is interesting to underline
that those works about subwavelength hole arrays have implications
in many other domains such as biosensor technologies or
left-handed metamaterials (see Ref. [16] for instance).
\\- The reader can refer to papers in Ref. [17] for plasmonic physics and technology
in the context introduced in the present paper.}

\end{document}